# Regioselective chemo-enzymatic syntheses of ferulate conjugates as chromogenic substrates for feruloyl esterases

Olga Gherbovet[1], Fernando Ferreira[1], Apolline Clément[1], Mélanie Ragon[1], Julien Durand[1], Sophie Bozonnet[1], Michael J. O'Donohue[1] and Régis Fauré*[1]

Address: [1]TBI, Université de Toulouse, CNRS, INRAE, INSA, Toulouse, France

Toulouse Biotechnology Institute, Bio & Chemical Engineering (TBI), Université de Toulouse - CNRS 5504 - INRAE 792 - INSA de Toulouse, 135 Avenue de Rangueil, 31077 Toulouse, France

Email: Régis Fauré - regis.faure@insa-toulouse.fr

* Corresponding author

# Abstract

Generally, carbohydrate-active enzymes are studied using chromogenic substrates that provide quick and easy color-based detection of enzyme-mediated hydrolysis. In the case of feruloyl esterases, commercially available chromogenic ferulate derivatives are both costly and limited in terms of their experimental application. In this study, we describe solutions for these two issues, using a chemoenzymatic approach to synthesize different ferulate compounds. The overall synthetic routes towards commercially available 5-bromo-4-chloro-3-indolyl and 4-nitrophenyl *O*-5-feruloyl-α-L-arabinofuranosides **1a** and **1b** were significantly shortened (7-8 steps reduced to 4-6) and transesterification yields enhanced (from 46 to 73% for **1a** and 47 to 86 % for **1b**). This was achieved using enzymatic (immobilized Lipolase 100T from *Thermomyces lanuginosus*) transesterification of unprotected vinyl ferulate to the primary hydroxyl group of α- L- arabinofuranosides. Moreover, a novel feruloylated-butanetriol 4-nitrocatechol-1-yl analog **12**, containing a cleavable hydroxylated linker was also synthesized in 29% overall yield in 3 steps (convergent synthesis). The latter route combined regioselective functionalization of 4-nitrocatechol and enzymatic transferuloylation. The use of **12** as a substrate to characterize type A feruloyl esterase from *Aspergillus niger* reveals the advantages of this substrate for the characterizations of feruloyl esterases.

## Keywords

# Introduction

The development of white biotechnology is underpinned by advances in enzyme discovery and engineering, areas that are being driven by metagenomics and *in vitro* directed enzyme evolution. These techniques procure massive discovery or creation of new enzyme-encoding sequences, filling up databases with a wealth of information. However, while resolving an early step in the discovery pipeline, these techniques progressively create a new bottleneck regarding enzyme characterization. Therefore, there is a pressing need to extend the enzymologist's toolbox, providing informationally rich high-throughput screens that can not only attribute an activity to putative enzymes, but also procure some qualitative details on enzyme properties. In this respect, the availability of easy to use chromogenic substrates that can provide both qualitative and quantitative assays and be compatible with automated protocols is a crucial issue. Feruloyl esterases (Faes; EC 3.1.1.73 and family CE1 of the CAZy classification [1]) are of interest, both because of their role in the deconstruction of complex plant-based materials and also as synthetic tools for the preparation of bioactive compounds with potential antioxidant properties [2–5]. Operating via a two-step serine protease mechanism involving a conserved Ser-His-Asp/Glu catalytic triad [6,7], Faes catalyse the hydrolysis of ester bonds linking hydroxycinnamoyl groups to the glycosyl moieties of plant-based polysaccharides, such arabinoxylans and arabinans. In this respect, Faes are important components of plant cell wall-degrading enzymatic arsenals, since the hydrolysis of *trans*-ferulate-polysaccharide linkages contributes to breakdown of intermolecular bonds that structure the lignocellulosic matrix. Moreover, Faes are useful tools to obtain commercially relevant ferulic acid, which represents up to 3% (w/w) of plant cell wall dry weight [8].

So far, the detection and characterization of Faes has mainly relied either on the use of HPLC or UV-visible spectrophotometry, using natural or synthetic compounds [9,10]. The latter, which are used in high-throughput screening (HTS) assays, fall into three categories. The simplest are feruloyl esters of chromogenic moieties [11–14], such as *p*-nitrophenol, or short chain alkyl groups (e.g., methyl ferulate). More elaborate and biologically relevant substrates contain a feruloylated L-arabinofuranosyl moiety [12,15–17]. These structurally more complex compounds are obtained using multi-step synthesis, considerably limiting availability. Moreover, they might be specific for certain subcategories of feruloyl esterases [18–20] and their use involves a tricky tandem reaction [21]. Finally, the synthesis of other more generic esters that can be used to assay esterases, including Faes, and lipases have been reported [22,23].

In this work, we revisit the preparation of simple feruloylated substrates, such as 5-bromo-4-chloro-3-indolyl and 4-nitrophenyl *O*-5-feruloylated α-L-arabinofuranosides **1a** and **1b**. Although these substrates are commercially available, their synthesis involves 7-8 steps [15–17]. This engenders rather high costs (i.e., as of July 29$^{th}$, 2020, €2500 and €778 per 100 mg for **1a** and **1b** respectively) that are approximately 19- and 14-fold higher than the non-feruloylated precursors. Therefore, our aim was to simplify synthesis in order to reduce cost. Furthermore, we describe the short synthesis of new feruloylated chromogenic substrate **12**, a molecule that obviates the need for a glycosyl moiety while containing a cleavable hydroxylated linker that mimics natural geometry and physico-chemical properties of osidic linkages.

## Results and Discussion

### Chemoenzymatic synthesis of 5-*O*-feruloylated α-L-arabinofuranosides

The synthesis of chromogenic 5-*O*-feruloylated α-L-arabinofuranosides **1a** and **1b** is usually achieved using a multi-step pathway that involves trapping the furanose conformation, anomeric activation, glycosidation, regioselective deprotection of the primary hydroxyl group, feruloylation and final deprotection to yield the target molecule [12,15–17]. Additionally, the temporary protection of functional groups is sometimes used during synthesis in order to facilitate certain steps. Using an alternative approach, we employed one-step regioselective transesterification of the unprotected vinyl ferulate **2** (synthesized in 56% *in-house* yield and up to 77% previously reported yield [24,25] in one-step and at gram scale) using immobilized Lipolase 100T from *T. lanuginosus* [26,27] and readily available and reasonably cheap 5-bromo-4-chloro-3-indolyl or 4-nitrophenylα-L-arabinofuranosides. This afforded the corresponding feruloylated derivatives, **1a** and **1b** (Figures 1A-B). Yields (73 and 86% for the indolyl and 4-nitrophenyl derivatives respectively) characterizing regioselective enzymatic feruloylation of the primary hydroxyl group compare favourably with previously reported overall yields (46 and 47% respectively, in three steps) [15–17,28], which relate to enzymatic selective *O*-5-deacetylation of the primary hydroxyl group, its esterification and final deprotection of the 2,3-*O*-acetyl groups of the glycoside and *O*-acetyl group of the ferulate moiety. That fact that lipase-catalysed transesterification obviates the need for require protection/deprotection is a considerable advantage, because the final deprotection in the chemical pathway is complicated by the presence of another ester linkage within the molecules [12–15]. In principle, the method described herein is generic and thus applicable to other chromogenic α-L-arabinofuranosidic compounds, such as 4-nitrocatechol (4NTC), 2-chloro-4-nitrophenyl and umbelliferyl derivatives.

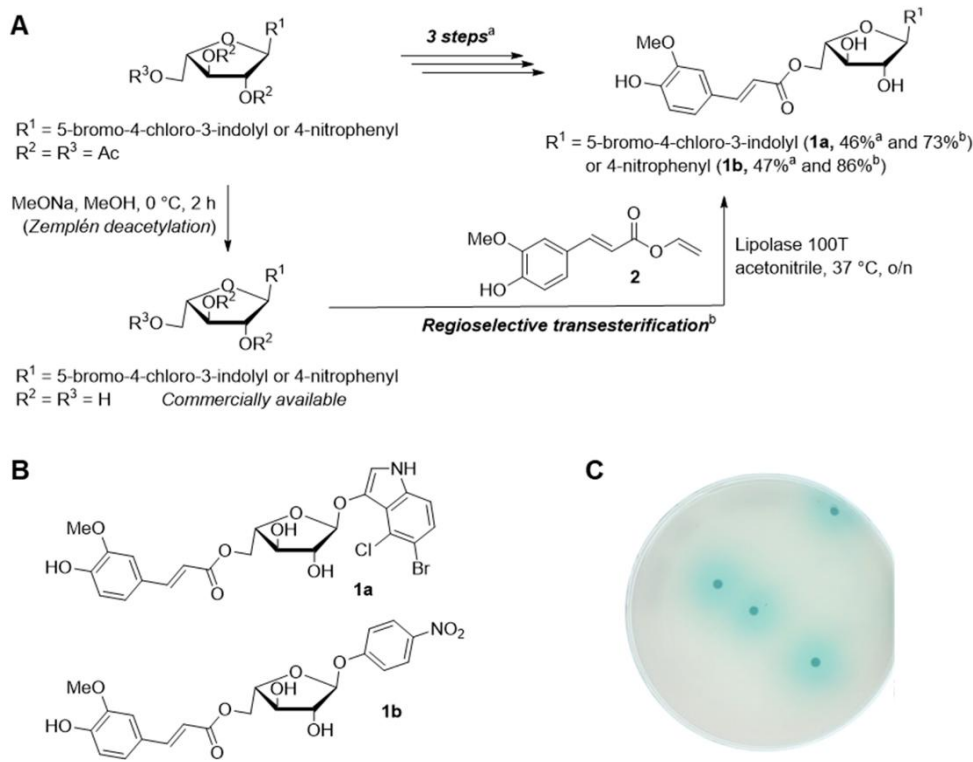

**Figure 1:** Alternative syntheses (A) and full structures (B) of 5-bromo-4-chloro-3-indolyl or 4-nitrophenyl 5-*O*-feruloyl-α-L-arabinofuranosidic chromogenic substrates, **1a** and **1b**, and (C) detection of type A Fae activity on solid agar medium using **1a**. The overall yields of transesterification procuring **1a** and **1b** are reported (in brackets) for both pathways ($3^a$ steps reduced to $1^b$).

To demonstrate the suitability of **1a** for qualitative *in situ* screening of microbial colonies growing on solid agar medium, the compound was incorporated (300 µg/mL) into top agar along with an α-L-arabinofuranosidase from *Thermobacillus xylanilyticus* (*Tx*Abf). This provided the means to reveal coloured microbial colonies expressing the type A Fae(+) from *Aspergillus niger* (*An*FaeA, Figure 1C). Coloration is the result of successive reactions: *(i)* release of the free 5-bromo-4-chloro-indoxyl-3-ol by an enzyme cascade, wherein *Tx*Abf-catalysed cleavage of the glycosidic bond was made possible by the prior release by *An*FaeA of the ferulate moiety [21]; and *(ii)* its spontaneous oxidation and subsequent dimerization forming the blue 5-bromo-4-chloro-indigo-like insoluble dye [15]. In an alternative demonstration, **1a** was also used for colony level detection of FAE using yeast cells (*Yarrowia lipolytica* [29]) that actually coexpressed *An*FaeA and *Tx*Abf(+) (data not shown).

## Investigating enzymatic transferuloylation reaction of non-glycosidic motifs

While lipase-catalysed transesterification reactions onto glycosidic structures have been extensively described, data related to the regioselectivity of transesterification on hydroxylated alkyl and/or aryl moieties are more sparse [30,31]. The extent to which Lipolase 100T catalyses feruloyl transfer reactions involving substituted benzylic alcohols was thus investigated to establish its usefulness for the preparation of various polyhydroxylated compounds of interest (e.g., antioxidants, chromogenic molecules for screening, etc.) [4,27]. Accordingly, we observed that transesterification only occurred when using 'primary' benzylic alcohols; no phenol acylation was detected in the case of hydroxynitrobenzylic alcohol (Table 1, no side-product of **4** and **9** with transfer on aromatic 'secondary' alcohol; i.e., phenol), 2-chloro-4-nitrophenol or 4NTC (data not shown). Additionally, the exact position of the benzyl alcohol affected transfer, with *ortho*-substitutions ($R^1$) displaying a hindered electron withdrawing group. This led to low target product yields (**5**, **6**, **8** and **9**), lower reactivity and/or greater hydrolysis of the vinyl ferulate into ferulic acid.

**Table 1:** Enzymatic transferuloylation of substituted nitro benzylic alcohols.

| R$^1$ | R$^2$ | R$^3$ | Product | Yield (%)$^a$ | Ratio$^b$ |
|---|---|---|---|---|---|
| H | H | NO$_2$ | 3 | 79 | 10/28/62 |
| OH | H | NO$_2$ | 4 | 76 | 19/32/49 |
| Cl | H | NO$_2$ | 5 | 44 | 31/38/31 |
| F | H | NO$_2$ | 6 | 46 | 33/28/39 |
| H | NO$_2$ | H | 7 | 97 | 10/24/66 |
| NO$_2$ | H | H | 8 | -$^c$ | 28/48/24 |
| NO$_2$ | H | OH | 9 | -$^c$ | 27/64/9 |

$^a$Yields of isolated ferulates after purification step; $^b$Ratio (in %), determined by $^1$H NMR, of the different species of feruloyl derivatives within the crude reaction mixture: remaining vinyl ferulate, ferulic acid (hydrolysis product) and ferulates **3-9**; $^c$The expected ferulates were confirmed by mass spectrometry analysis (HRMS) but the low purity of samples after purification prevented fine structural characterization by NMR.

## Synthesis of L-arabinofuranoside free 4-nitrocatechol-1-yl-linker-ferulate chromogenic substrate (12) and its evaluation as chromogenic substrate for Fae assays

To synthesize the chromogenic feruloylated-butanetriol 4-nitrocatechol-1-yl **12** (4NTC-linker-Fe), which contains 4NTC bound via a cleavable linker to a ferulate motif, a multi-step route was devised (Scheme 1). First, a new shorter, more practical pathway towards (*S*)-4-*O*-(2-hydroxy-4-nitrophenyl)-1,2,4-butanetriol **11** was developed. Compared to the previously reported 4-step synthesis [32], two drawbacks were circumvented, notably avoiding *(i)* the preparation of the volatile (*S*)-1-iodo-3,4-*O*-isopropylidene-3,4-butanediol intermediate and *(ii)* the use of a protected version of the chromogenic linker **11**, either for the extra hydroxyl group of the catecholyl moiety or the secondary hydroxyl group of the linker [23,32]. Alkylation of 4NTC with homoallylic bromide in basic conditions gave a mixture of mono- and di-alkylated 4NTC derivatives with **10** (38%) predominating because of the preferential formation of the phenolate at the *para* position (relative to the nitro group) [33]. Osmium tetroxide-mediated dihydroxylation in the presence of *N*-methylmorpholine *N*-oxide (NMMO) afforded **11** in 81% yield. Finally, regioselective transferuloylation of the primary hydroxyl of triol derivative **11** with Lipolase 100T was performed and the expected chromogenic substrate **12** was isolated in high 94% yield. Accordingly, the synthesis of chromogenic ferulate **12** was achieved in 29% overall yield in 3 steps from commercial reactants (convergent synthesis using a slight excess of synthesized vinyl ferulate **2**) and without the requirement to perform final deprotection.

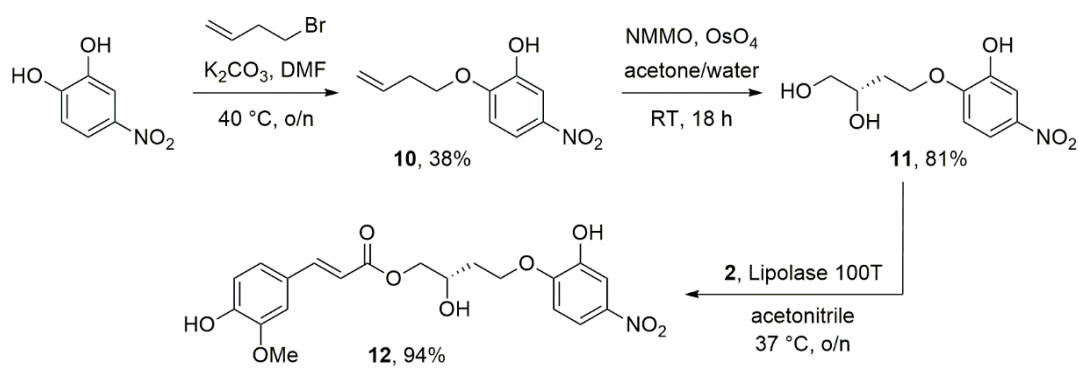

**Scheme 1:** Chemoenzymatic synthesis of (*S*)-4-*O*-(2-hydroxy-4-nitrophenyl)-1-*O*-*trans*-feruloyl-1,2,4-butanetriol (4NTC-linker-Fe, **12**).

As expected, investigation of the stability of the chromogenic substrate **12** using UV-visible spectrophotometry revealed that, unlike 4-nitrocatechol-1-yl ferulate (4NTC-Fe) that undergoes spontaneous hydrolysis even at neutral pH and 40 °C [13], the presence of the alkyl-like linker procures higher stability over a wide pH range (up to pH 9.0), irrespective of temperature. This is because in compound **12** the ferulate moiety is not directly linked to the good leaving group 4NTC ($pK_a$ = 6.61 [34]). Instead it is bonded to the linker whose $pK_a$ can be assimilated either with that of glycerol ($pK_a$ = 13.61) or L-arabinose with ($pK_a$ =11.31) [35], meaning that it is a poor leaving group. Moreover, our observations regarding linker stability are consistent with the known stability of ester linkages under basic conditions.

The usefulness of 4NTC-linker-Fe **12** for the characterization of Faes was investigated (Figure 2), measuring 4NTC release by *An*FaeA [36] at 40 °C. The enzyme-catalysed reaction leads to cleavage of the ester bond linking the ferulate to the linker-4NTC moiety and thus accumulation of linker-NTC. Therefore, working in discontinuous mode, 4NTC is quantified by submitting samples removed from the reaction to the oxidative action of sodium periodate at 0°C and reading absorbance at 530 nm (in alkaline conditions) [23,32,37,38]. Importantly, it is vital to include a stoichiometric amount of ethylene glycol to avoid further oxidation of free NTC by sodium periodate (Figure 2B).

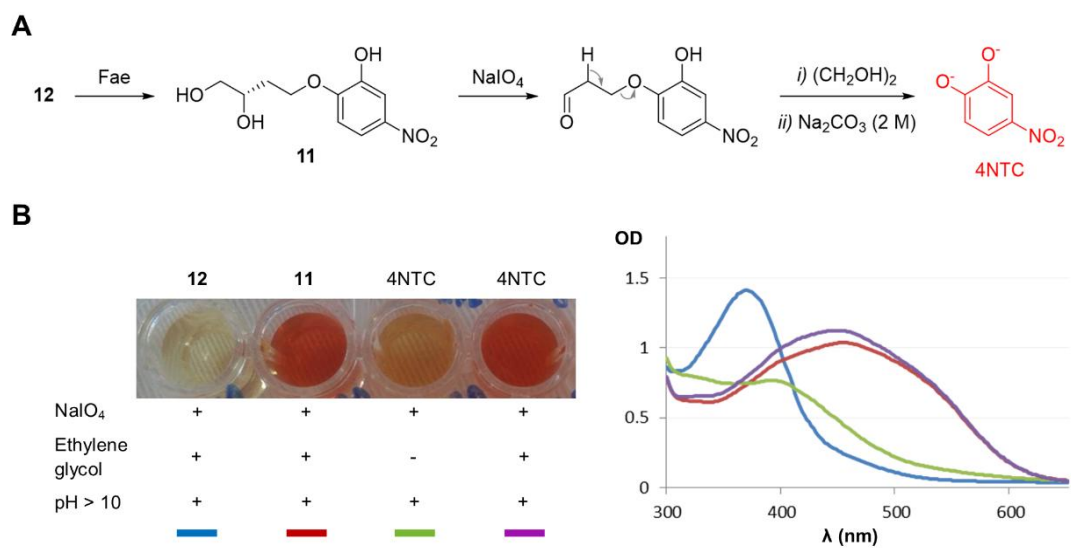

**Figure 2:** (A) Principle of the 4NTC release after action of Fae on **12** in the presence of sodium periodate and (B) control reaction.

The specific activity (SA) of *An*FaeA on 4NTC-linker-Fe **12** was determined to be 3 IU/mg of protein, a value comparable to that measured on destarched wheat bran (3 IU/mg) [39] containing 5-*O*-feruloylated α-L-arabinofuranosyl moieties, but lower than that (40 IU/mg, unpublished data) measured using more labile 4NTC-Fe. Therefore, although 4NTC-Fe is a practical synthetic probe for both high-throughput screening and preliminary characterisation of Fae activity [13], 4NTC-linker-Fe **12** is almost certainly a better analogue of ferulate linkages found in plant-based structures.

## Conclusion

The use of immobilized Lipolase 100T provides the means to perform regioselective transesterification of vinyl ferulate **2** to the primary hydroxyl group of benzylic alcohols and polyhydroxylated compounds. Three compounds suitable for the detection and/or characterization of Fae activity were synthesized in a straightforward protocol that holds the potential to greatly reduce the cost of substrates **1a** and **1b**. Moreover, the enzyme-driven convergent synthesis of **12** affords a novel substrate that is highly suitable for the characterization of feruloyl esterases.

# Experimental

## Materials and general methods

4-Nitrophenyl and 5-bromo-4-chloro-3-indolyl α-L-arabinofuranosides were purchased from Carbosynth (Compton, U.K.) and immobilized Lipolase 100T (known as Lipozyme® TL IM, from *T. lanuginosus*; 250 IUN/g with IUN = interesterification unit) was supplied by Novozymes (Bagsvaerd, Denmark). Reaction evolution was monitored by analytical thin-layer chromatography using silica gel 60 F254 precoated plates (E. Merck). Spots were visualized using UV light of 254 nm wavelength followed by soaking in a 0.1% (w/v) orcinol solution containing a mixture of sulfuric acid/ethanol/water (3:72.5:22.5 v/v/v) followed by charring. Purifications by column chromatography were performed using a Reveleris® flash chromatography automated system (BUCHI, Villebon-sur-Yvette, France) equipped with prepacked irregular silica gel 40-63 μm cartridges (FlashPure EcoFlex, BUCHI). NMR spectra were recorded on a Bruker Avance II 500 spectrometer at 298 K. Chemical shifts ($\delta$) are given in ppm with residual solvents signal as internal reference [40]. Coupling constants ($J$) are reported in Hertz (Hz) with singlet (s), doublet (d), triplet (t), doublet of doublet (dd), doublet of doublet of doublets (ddd), broad (br) and quadruplet of triplet (qt). Analysis and assignments were made using 1D ($^1$H, $^{13}$C and J-modulated spin-echo ($J_{mod}$)) and 2D (COrrelated SpectroscopY (COSY) and Heteronuclear Single Quantum Coherence

(HSQC)) experiments. High-resolution mass spectra (HRMS) analyses were performed at PCN-ICMG (Grenoble, France).

**General procedure for enzymatic transesterification**

The enzymatic transesterification steps were performed according to a published protocol [26]. Briefly, Lipolase 100T (1 g) was added to a solution of alcohol (0.30 mmol, 1 equiv.) and vinyl ferulate **2** (100 mg, 0.45 mmol, 1.5 equiv.) in acetonitrile (6 mL). The reaction mixture was stirred overnight at 37 °C, then filtered, filter cake washed with acetone and the filtrate was evaporated to dryness. The residue was recovered in ethyl acetate, washed with saturated aqueous sodium hydrogencarbonate (three times). Combined organic phases were dried over anhydrous sodium sulfate, filtered and concentrated under reduced pressure. Flash chromatography (gradient of ethyl acetate in petroleum ether from 0 to 50%) afforded pure ferulates **1a**, **1b**, **3-9** and **12**.

**3-Nitrobenzyl *trans*-ferulate** (**3**, 78 mg. 0.24 mmol, 79%), white powder. $^1$H NMR (500 MHz, (CD$_3$)$_2$CO): $\delta$ 8.32 (1H, t, *J* 1.7, CH of Bn), 8.21 (1H, ddd, *J* 8.3, 2.2, 1.1, CH of Bn), 7.90 (1H, ddd, *J* 7.6, 1.7, 1.0, CH of Bn), 7.71 (1H, t, *J* 7.6, CH of Bn), 7.68 (1H, d, *J* 15.8, CH=CHCO$_2$), 7.37 (1H, d, *J* 2.0, CH of Fe), 7.17 (1H, dd, *J* 8.3, 2.0, CH of Fe), 6.87 (1H, d, *J* 8.3, CH of Fe), 6.50 (1H, d, *J* 15.8, CH=CHCO$_2$), 5.38 (2H, s, CH$_2$ of Bn), 3.92 (3H, s, OMe). $^{13}$C NMR (125 MHz, (CD$_3$)$_2$CO): $\delta$ 167.2 (C=O), 150.3 (C$_q$), 149.4 (C$_q$), 148.8 (C$_q$) 146.7 (CH=CHCO$_2$), 140.2 (C$_q$), 135.1 (CH of Bn), 130.8 (CH of Bn), 127.3 (2C$_q$), 124.2 (CH of Fe), 123.6 (CH of Bn), 123.4 (CH of Bn), 116.1 (CH of Bn), 115.1 (CH=CHCO$_2$), 111.4 (CH of Fe), 65.1 (CH$_2$ of Bn), 56.4 (OMe). HRMS (ESI) calc. for [M-H]$^-$ C$_{17}$H$_{14}$NO$_6$ *m/z* 328.0821, found: 328.0833.

**2-Hydroxy-5-nitrobenzyl *trans*-ferulate** (**4**, 79 mg, 0.23 mmol, 76%), white powder. $^1$H NMR (500 MHz, (CD$_3$)$_2$CO): δ 8.29 (1H, d, *J* 2.8, CH of Bn), 8.14 (1H, dd, *J* 8.9, 2.8, CH of Bn), 7.68 (1H, d, *J* 15.8, CH=CHCO$_2$), 7.38 (1H, d, *J* 1.8, CH of Fe), 7.17 (1H, dd, *J* 8.0, 1.8, CH of Fe), 7.12 (1H, d, *J* 8.9, CH of Bn), 6.88 (1H, d, *J* 8.0, CH of Fe), 6.52 (1H, d, *J* 15.8, CH=CHCO$_2$), 5.33 (2H, s, CH$_2$ of Bn), 3.92 (3H, s, OMe). $^{13}$C NMR (125 MHz, (CD$_3$)$_2$CO): δ 167.5 (C=O), 162.0 (C$_q$) 150.3 (C$_q$), 148.8 (C$_q$), 148.7 (C$_q$), 146.6 (CH=CHCO$_2$), 141.6 (C$_q$), 127.4 (C$_q$), 126.3 (2CH of Bn), 124.3 (CH of Fe), 116.5 (CH of Bn), 116.1 (CH of Fe), 115.2 (CH=CHCO$_2$), 111.3 (CH of Fe), 61.1 (CH$_2$ of Bn), 56.4 (OMe). HRMS (ESI) calc. for [M-H]$^-$ C$_{17}$H$_{14}$NO$_7$ *m/z* 344.0770, found: 344.0778.

**2-Chloro-5-nitrobenzyl *trans*-ferulate** (**5**, 48 mg, 0.13 mmol, 44%), white powder. $^1$H NMR (500 MHz, (CD$_3$)$_2$CO): δ 8.42 (1H, br d, *J* 3.0, CH of Bn), 8.26 (1H, dd, *J* 9.0, 3.0, CH of Bn), 7.81 (1H, d, *J* 9.0, CH of Bn), 7.71 (1H, d, *J* 16.0, CH=CHCO$_2$), 7.40 (1H, d, *J* 1.8, CH of Fe), 7.19 (1H, dd, *J* 8.0, 1.8, CH of Fe), 6.88 (1H, d, *J* 8.0, CH of Fe), 6.56 (1H, d, *J* 16.0, CH=CHCO$_2$), 5.43 (2H, s, CH$_2$ of Bn), 3.93 (3H, s, OMe). $^{13}$C NMR (125 MHz, (CD$_3$)$_2$CO): δ 167.1 (C=O), 164.6 (C$_q$), 150.4 (C$_q$), 148.8 (C$_q$), 147.1 (CH=CHCO$_2$), 140.6 (C$_q$), 137.4 (C$_q$), 131.7 (CH of Bn), 127.3 (C$_q$), 125.2 (2CH of Bn), 124.4 (CH of Fe), 116.1 (CH of Fe), 114.7 (CH=CHCO$_2$), 111.4 (CH of Fe), 63.0 (CH$_2$ of Bn), 56.4 (OMe). HRMS (ESI) calc. for [M-H]$^-$ C$_{17}$H$_{13}$ClNO$_6$ *m/z* 362.0431, found: 362.0431.

**2-Fluoro-5-nitrobenzyl *trans*-ferulate** (**6**, 48 mg, 0.14 mmol, 46%), white powder. $^1$H NMR (500 MHz, (CD$_3$)$_2$CO): δ 8.43 (1H, dd, *J* 3.2, 2.9, CH of Bn), 8.33 (1H, ddd, *J* 9.0, 4.3, 3.0 Hz, CH of Bn), 7.67 (1H, d, *J* 16.0, CH=CHCO$_2$), 7.49 (1H, t, *J* 9.0, CH of Bn), 7.36 (1H, d, *J* 1.8, CH of Fe), 7.16 (1H, dd, *J* 8.0, 1.8, CH of Fe), 6.87 (1H, d, *J*

8.0, CH of Fe), 6.49 (1H, d, *J* 16.0, CH=CHCO$_2$), 5.39 (2H, s, CH$_2$ of Bn), 3.91 (3H, s, OMe). $^{13}$C NMR (125 MHz, (CD$_3$)$_2$CO): δ 167.1 (C=O), 165.1 (d, *J* 257.4, C$_q$), 150.4 (C$_q$), 148.8 (C$_q$), 146.9 (CH=CHCO$_2$), 127.2 (C$_q$), 127.1 (d, *J* 6.1, CH of Bn), 126.9 (d, *J* 10.5, CH of Bn), 126.7 (d, *J* 17.1, C$_q$), 124.3 (CH of Fe), 117.7 (d, *J* 24.3, CH of Bn), 116.1 (CH of Fe), 114.7 (CH=CHCO$_2$), 111.4 (CH of Fe), 59.6 (d, *J* 3.9, CH$_2$ of Bn), 56.3 (OMe). HRMS (ESI) calc. for [M-H]$^-$ C$_{17}$H$_{13}$FNO$_6$ *m/z* 346.0727, found: 346.0731.

**4-Nitrobenzyl *trans*-ferulate** (**7**, 96 mg, 0.29 mmol, 97%), white powder. $^1$H NMR (500 MHz, (CD$_3$)$_2$O): δ 8.27 (2H, m, CH of Bn), 7.72 (2H, m, CH of Bn), 7.69 (1H, d, *J* 16.0, CH=CHCO$_2$), 7.35 (1H, d, *J* 2.0, CH of Fe), 7.17 (1H, dd, *J* 8.2, 2.0, CH of Fe), 6.88 (1H, d, *J* 8.2, CH of Fe), 6.51 (1H, d, *J* 16.0, CH=CHCO$_2$), 5.38 (2H, s, CH$_2$ of Bn), 3.92 (3H, s, OMe). $^{13}$C NMR (125 MHz, (CD$_3$)$_2$CO): δ 167.1 (C=O), 150.3 (C$_q$), 148.8 (C$_q$), 148.6 (C$_q$) 146.7 (CH=CHCO$_2$), 145.4 (C$_q$), 129.4 (CH of Bn), 127.3 (C$_q$), 124.4 (CH of Fe), 124.2 (CH of Bn), 116.0 (CH of Fe), 115.0 (CH=CHCO$_2$), 111.4 (CH of Fe), 65.1 (CH$_2$ of Bn), 56.4 (OMe). HRMS (ESI) calc. for [M-H]$^-$ C$_{17}$H$_{14}$NO$_6$ *m/z* 328.0821, found: 328.0825.

## Synthesis of feruloylated-butanetriol 4-nitrocatechol-1-yl (4NTC-linker-Fe, 12)

**2-*O*-(But-3-enyloxy)-5-nitrophenol** (**10**). To a solution of 4-nitrocatechol (4NTC; 1.00 g, 6.44 mmol, 1 equiv.) in dry DMF (8 mL) were added potassium carbonate (1.06 g, 7.67 mmol, 1.2 equiv.) and homoallylic bromide (670 μL, 6.51 mmol, 1 equiv.) at 40 °C. After overnight stirring at 40 °C, the reaction mixture was concentrated under reduced pressure. The residue was recovered in ethyl acetate, washed with saturated aqueous sodium hydrogencarbonate and brine. Combined organic phases were dried over anhydrous sodium sulfate, filtered and concentrated under reduced pressure. Flash chromatography (gradient of ethyl acetate in petroleum ether from 0 to 50%) afforded **10** (516 mg, 2.47 mmol, 38%), as a solid. 1H NMR (500 MHz, CDCl3): $\delta$ 7.82 (1H, dd, *J* 9.0, 3.0, CH of 4NTC), 7.80 (1H, d, *J* 3.0, CH of 4NTC), 6.90 (1H, d, *J* 9.0, CH of 4NTC), 5.91-5.83 (1H, m, =CH), 5.23-5.16 (2H, m, $CH_2$=), 4.21 (2H, t, *J* 6.6, $CH_2O$), 2.63 (2H, br qt, *J* 6.6, 1.6, $CH_2$). $^{13}$C NMR (500 MHz, CDCl$_3$): $\delta$ 151.3 ($C_q$), 145.8 ($C_q$), 142.1($C_q$), 133.2 (=CH), 118.1 ($CH_2$=), 116.8 (CH of 4NTC), 110.4 (CH of 4NTC), 110.1 (CH of 4NTC), 68.5 ($CH_2O$), 33.2 ($CH_2$). HRMS (ESI) calc. for [M-H]$^-$ $C_{10}H_{10}NO_4$ *m/z* 208.0610, found: 208.0615.

(***S***)-**4-*O*-(2-Hydroxy-4-nitrophenyl)-1,2,4-butanetriol** (**11**, [32]). A solution of **10** (157 mg, 0.75 mmol, 1 equiv.) in an acetone/water mixture (2.5-1 v/v, 3.5 mL) was treated at 25 °C under stirring with *N*-methylmorpholine-*N*-oxide (NMMO; 106 mg, 0.90 mmol, 1.2 equiv.) and osmium tetroxide (2.5 wt % solution in *tert*-butanol, 38 μL) and stirred at room temperature for 18 h. 10% (w/v) Aqueous sodium sulfite (0.5 mL) was added and stirring was prolonged for 30 min. The product was extracted with ethyl acetate (three times with 10 mL), washed with brine. Combined organic phases, dried over anhydrous sodium sulfate, filtered and concentrated under reduced pressure, afforded the expected compound **11** (165 mg, 0.68 mmol, 90%) as a solid.

(***S***)-**4-*O*-(2-Hydroxy-4-nitrophenyl)-1-*O*-*trans*-feruloyl-1,2,4-butanetriol** (**12**). Application of the general procedure for enzymatic transesterification with **11** (70 mg, 0.29, mmol, 1 equiv.) and vinyl ferulate **2** (94 mg, 0.43 mmol, 1.5 equiv.) to give **12** (113 mg, 0.27 mmol, 94%) as a white powder. $^1$H NMR (500 MHz, (CD$_3$)$_2$CO): $\delta$ 7.79 (1H, dd, *J* 9.0, 2.8, CH of 4NTC ), 7.67 (1H, d, *J* 2.8, CH of 4NTC), 7.63 (1H, d, *J* 16.0, CH=CHCO$_2$), 7.33 (1H, d, *J* 2.0, CH of Fe), 7.20 (1H, d, *J* 9.0, CH of 4NTC), 7.13 (1H, dd, *J* 8.2, 2.0, CH of Fe), 6.87 (1H, d, *J* 8.2, CH of Fe), 6.40 (1H, d, *J* 16.0, CH=CHCO$_2$), 4.47-4.42 (1H, m), 4.40-4.36 (1H, m), 4.23-4.16 (3H, m), 3.95 (3H, s, OMe), 2.18-2.11 (1H, m), 2.01-1.94 (1H, m). $^{13}$C NMR (125 MHz, (CD$_3$)$_2$CO): δ 167.4 (C=O), 153.6 (C$_q$), 150.2 (C$_q$), 148.8 (C$_q$), 147.7 (C$_q$), 146.0 (CH=CHCO$_2$), 142.4 (C$_q$), 127.4 (C$_q$), 124.0 (CH of Fe), 117.1 (CH of 4NTC), 116.1 (CH of Fe), 115.7 (CH=CHCO$_2$), 112.4 (CH of 4NTC), 111.3 (CH of Fe), 110.8 (CH of 4NTC), 69.0 (CH$_2$), 67.1 (CH$_2$), 66.9 (CH), 56.3 (OMe), 33.7 (CH$_2$). HRMS (ESI) calc. for [M-H]$^-$ C$_{20}$H$_{20}$NO$_9$ *m/z* 418.1138, found: 418.1139.

## Screening of Fae(+) microorganisms in solid medium using X-α-L-Ara*f*-Fe (1b)

*Y. lipolytica An*FaeA(+) strain was used to inoculate solid YNB medium (1.7 g/L YNB without casamino acid, 5 g/L ammonium chloride, 20 mL/L oleic acid, 10 g/L D-glucose, 2 g/L casamino acid, and 15 g/L bacto agar in 100 mM citrate-phosphate buffer pH 5. Petri dishes were incubated for 48 h at 30 °C and then overlayed with a preparation of molten 1% (w/v) top agar containing chromogenic substrate **1a** (300 µg/mL and 0.5% DMSO) and *Tx*Abf (2 IU/mL). Once the top agar was solid, incubation at 37 °C for 1 h allows color to develop. *Y. lipolytica* strains that contain no *An*FaeA gene were also checked to remain colorless after addition of chromogenic substrate and auxiliary enzyme.

## Liquid medium-based colorimetric assays using 4NTC-linker-Fe (12)

In a typical experiment, discontinuous enzyme assays were performed in triplicate in buffered conditions (100 mM sodium phosphate pH 6.0) in the presence of 1.8 mM **12** and 3.6% DMSO, final concentrations. For the assay, this solution was preincubated at 40 °C before *An*FaeA addition. Aliquots (25 µL) were stopped by cooling (at 0 °C) every 6 min over a 24 min-period and mixed with 45 µL of cooled 10 mM $NaIO_4$ solution (pH 2.0). After keeping 5 min at 0 °C throughout, 45 µL of ethylene glycol were added, followed by 135 µL of 2 M $Na_2CO_3$ after 5 min. The optical densities (OD) at 530 nm were recorded on a microplate reader Infinite M200 PRO (TECAN). One unit (IU) of Fae specific activity (SA, expressed in µmol/min/mg or IU/mg) corresponds to the amount of released 4NTC (in µmol) per minute per milligram of protein. Negative controls containing all of the reactants except the enzyme were always included in order to monitor and correct for spontaneous hydrolysis of the substrate. Control reactions containing **12**, **11** and 4NTC without enzyme and 4NTC without both enzyme and ethylene glycol were also prepared.

## Acknowledgments

The NMR work carried out in this work at TBI (Toulouse, France) was performed with the equipment of Meta-Toul (Metabolomics & Fluxomics Facitilies, Toulouse, France, www.metatoul.fr). MetaToul is part of the national infrastructure MetaboHUB (The French National infrastructure for metabolomicsand fluxomics, www.metabohub.fr) and is supported by grants from the Région Midi-Pyrénées, the European Regional Development Fund, SICOVAL, IBiSa-France, CNRS and INRAE. We thank the ICEO facility dedicated to enzyme screening and discovery, and part of the Integrated Screening Platform of Toulouse (PICT, IBiSA) for providing access to its equipment. The authors wish to acknowledge the support from the ICMG Chemistry Nanobio Platform (Grenoble, France) for HRMS analyses.

# Funding

This work was supported by the European Union's Seventh Programme for Research, Technological Development and Demonstration under Grant Agreement No 613868, OPTIBIOCAT project (to O.G. and J.D.), by the French National Research Agency, grant ANR-05-PNRB-002, project SPPECABBE (to M.R.) and the Région Midi-Pyrénées grants DAER-Recherche 07009817(to F.F.).

# ORCID® iDs

Apolline Clément - https://orcid.org/0000-0002-8651-7566

Mélanie Ragon - https://orcid.org/0000-0002-1676-4764

Julien Durand - https://orcid.org/0000-0002-5631-6210

Sophie Bozonnet - https://orcid.org/0000-0001-5091-2209

Michael J. O'Donohue - https://orcid.org/0000-0003-4246-3938

Régis Fauré - https://orcid.org/0000-0002-5107-9009